\documentclass[a4paper,10pt]{scrartcl}

\usepackage{amsmath}
\usepackage{amsfonts}
\usepackage{amssymb}
\usepackage{graphicx}
\usepackage[english]{babel}
\usepackage{wasysym}
\usepackage{units}

\newcommand{\vs}[1]{\vspace{#1mm}}
\newcommand{\vsO}{\vspace{.1cm}\hfill\\}
\newcommand{\vsT}{\vspace{.2cm}\hfill\\}

\title{\Large Macroscopic and Microscopic Paradigms for the Torsion Field: from the Test-Particle Motion to a Lorentz Gauge Theory}

\author{{\large N. Carlevaro}$^{\;a,\,b}$, {\large O.M. Lecian}$^{\;a,\,c}$ 
{\large and G. Montani}$^{\;a,\,c,\,d,\,e}$\vsT
\emph{\footnotesize $^a$ICRA -- International Center for Relativistic Astrophysics,}\vs{-2.5}\\
\emph{\footnotesize c/o Dep. of Physics - ``Sapienza'' Universit\`a di Roma}\\
\emph{\footnotesize $^b$Department of Physics, Polo Scientifico -- Universit\`a degli Studi di Firenze,}\vs{-2.5}\\
\emph{\footnotesize INFN -- Section of Florence, Via G. Sansone, 1 (50019), Sesto Fiorentino (FI), Italy}\\
\emph{\footnotesize $^c$ Department of Physics - ``Sapienza'' Universit\`a di Roma, Piazza A. Moro, 5 (00185), Rome, Italy}\\
\emph{\footnotesize $^d$ENEA -- C.R. Frascati (Department F.P.N.), Via Enrico Fermi, 45 (00044), Frascati (Rome), Italy}\\
\emph{\footnotesize $^{e}$ ICRANet -- C. C. Pescara, Piazzale della Repubblica, 10 (65100), Pescara, Italy}\\
{\footnotesize\ttfamily nakia.carlevaro@icra.it\quad lecian@icra.it\quad montani@icra.it}
}
\date{}
\begin{document}
\maketitle

%
\begin{abstract} \textbf{Abstract:} Torsion represents the most natural extension of General Relativity and it attracted interest over the years in view of its link with fundamental properties of particle motion. The bulk of the approaches concerning the torsion dynamics focus their attention on their geometrical nature and they are naturally led to formulate a non-propagating theory.

Here we review two different paradigms to describe the role of the torsion field, as far as a propagating feature of the resulting dynamics is concerned. However, these two proposals deal with different pictures, \emph{i.e.},  a macroscopic approach, based on the construction of suitable potentials for the torsion field, and a microscopic approach, which relies on the identification of torsion with the gauge field associated with the local Lorentz symmetry. We analyze in some detail both points of view and their implications on the coupling between torsion and matter will be investigated. In particular, in the macroscopic case, we analyze the test-particle motion to fix the physical trajectory, while, in the microscopic approach, a natural coupling between torsion and the spin momentum of matter fields arises. 

\vsO \emph{PACS}: 02.40.-k; 04.20.Fy; 04.50.+h; 11.15.-q
\end{abstract}

\section{Introduction}
Completely neglected in the first formulation of the \emph{General Relativity} (GR), \emph{torsion} was later taken into consideration principally by \'E. Cartan \cite{crtn1}. The usual version of Einstain-Cartan theory \cite{hhl1,sciama} is based on the standard Einstein action, where the scalar curvature is a function of both metric and torsion. From variational principles, field equations are obtained in presence of matter, and it can be pointed out that, in such a theory, torsion is not really a dynamical field in the same sense as the metric field. From a microscopic point of view, recent studies on the coupling of torsion with spinor matter are those in \cite{hhl1,sciama,uti56,kib61,hehl74,deis} and \cite{blago,blago2}. In the $U_4$ theory \cite{hhl1,sciama,uti56,kib61,hehl74,deis}, torsion corresponds to the rotation gauge potential, and it is related to the intrinsic angular momentum of matter. In Poincar\'e Guage Theory (PGT) \cite{blago,blago2}, torsion and bein vectors are the gauge fields that account for local Poincar\'e transformations. These two descriptions predict a non-propagating torsion field, so that only a contact interaction is expected, because the equations of motion are algebraic rather than differential.\\
Contrastingly, in this paper, we will propose microscopic and macroscopic approaches, which predict a propagating torsion field. In both these schemes, the dynamics of torsion will acquire particular features that imply interesting perspectives about it detection.\\
The paper is organized as follows.\\
In Sec. \textbf{2}, the macroscopic approach is developed by some assumptions about the form of the torsion tensor \cite{amr03}: the completely antisymmetric and trace part of the tensor are considered derived from two local torsion potential. Then, by the action principle, we determine the field equations for these potentials, which are wave equations ideed. The motion equation of test particles are determined as autoparallels and the non-relativistic limit of these trajectories and of the tidal effects show that the torsion trace potential $\phi$ enters all the equations in the same way as the gravitational potential. In Sec. \textbf{3}, propagating torsion will be also derived form a microscopic point of view \cite{lmm,lm07}. In fact, the introduction of a Lorentz gauge field on flat space-time will allow us to identify the Lorentz gauge field with torsion, and, on curved space-time, all the geometric features of this interaction will be investigated. The comparison of first- and second-order approaches will be explained in the linearized regime, where the role of the gravitational field as a source for torsion will be compared with the spin-current term of the second-order formalism. Concluding remarks follow.

\section{Macroscopic paradigm: test-particle motion}

\subsection{Lagrangian geometric theory with propagating torsion}

In non-flat spaces, the concept of parallel transport of vector fields needs the introduction of connections, which also define the covariant derivative. The usual contruction of such a derivative (denoted by $\nabla_\mu$)
\footnote{Greek indices $\mu=0,1,2,3$ transform under general coordinate transformations, while Latin indices from the middle of the alphabet takes the values $i=1,2,3$.}
is performed by means of the affine-connection coefficients $\Gamma_{\mu\nu}^\rho$\,, which are, in general, non-tensor quantities. On the other hand, their pure antisymmetric part, called \emph{torsion} $\;T_{\mu\nu}^{\rho}=\Gamma_{[\mu\nu]}^\rho$, transforms like a tensor, as fas as the most general metric-compatible form of connections are concerned. The introduction of torsion was due principally by \'E. Cartan \cite{crtn1}, according to whom torsion was connected with intrinsic angular momentum. Later, this idea was extended by F. Hehl et al. \cite{hhl1}, which identified torsion to the rotation gauge potential.

We now introduce a metric $g_{\mu\nu}$ in an Einstein-Cartan space $U_4$ and require that the non-metricity $Q_{\mu\nu\rho}=-\nabla_\mu g_{\nu\rho}$ be vanishing. In this picture, connection coefficients write as
\begin{equation}\label{connections_contortion}
\Gamma_{\mu\nu\rho}=\tilde{\Gamma}_{\mu\nu\rho}+ K_{\mu\nu\rho}\;,\qquad
K_{\mu\nu\rho}=
\nicefrac{1}{2}[\,T_{\mu\nu\rho}-T_{\nu\rho\mu}+T_{\rho\mu\nu}\,]\;,
\end{equation}
where $\tilde{\Gamma}_{\mu\nu\rho}$ are the usual Christoffel symbols  (the symbol $({\scriptscriptstyle{\sim}})$ stands for Riemannian) and $K_{\mu\nu\rho}$ identifies the \emph{contortion} tensor.  

\paragraph{The torsion tensor} Torsion is a three-index tensor, antisymmetric in the first two indices; according to group theory, it can be decomposed in a completely antisymmetric part, a trace part and a third part with no special symmetry properties. In our analysis, we consider only the first two terms and we assume they to be derived from the exterior derivative of two \emph{potentials},
\begin{equation}\label{potentials}
B_{\mu\nu\rho}\equiv T_{[\mu\nu\rho]}=\tilde{\nabla}_{[\mu}A_{\nu\rho]}\;,\qquad
Tr{\scriptstyle{[\,T_{\mu\nu\rho}\,]}}=
\nicefrac{1}{3}(g_{\nu\rho}\partial_\mu\phi-g_{\mu\rho}\partial_\nu\phi)\;,  
\end{equation}
where $A_{\mu\nu}(x)$ is an antisymmetric tensor, while $\phi(x)$ is a scalar. This way contortion writes as $K_{\mu\nu\rho}=B_{\mu\nu\rho}+2\,Tr{\scriptstyle{[\,T_{\mu\nu\rho}\,]}}$. The introduction of potentials for the antisymmetric part of torsion \cite{hmmnd,hjmn,nomura} has its main motivation just in obtaining a propagating field in vacuum.

\paragraph{Field equations} To calculate field equations, we now introduce the usual Hilbert-Einstein action, which can be split up in its Riemannian part plus torsion-depending terms:
\begin{align}\label{action}
\mathcal{S}_{HE}=-\nicefrac{1}{2k}\textstyle{\int} dx\sqrt{-g}\;
(\tilde{R}-B^{\mu\nu\rho}B_{\mu\nu\rho}-\nicefrac{2}{3}(\partial_\mu\phi)^2)\;.
\end{align}
We obtain field equations by variational principles: variations with respect to $g_{\mu\nu}$, $A_{\mu\nu}$ and $\phi$ yield, respectively,
\begin{align}
-&\tilde{G}^{\mu\nu}-\nicefrac{1}{2}\;g^{\mu\nu}B^{\rho\sigma\epsilon}B_{\rho\sigma\epsilon}+
3B^{\mu\sigma\epsilon}B^{\nu}_{\sigma\epsilon}-\nicefrac{8}{3}\;(\nicefrac{1}{2}\;g^{\mu\nu}(\partial_\rho\phi)^2-
g^{\mu\rho}g^{\nu\sigma}(\partial_\rho\phi)(\partial_\sigma\phi))=0\;,\label{e:3_1}\\
&\tilde{\nabla}_\mu B^{\mu\nu\rho}=0\;,  \label{e:3_2}\\
&\tilde{\nabla}_\mu g^{\mu\nu}\partial_\nu\phi=0. \label{e:3_3}
\end{align}
Eq. (\ref{e:3_1}) consists of the (Riemannian) Einstein tensor, as in GR, plus four terms all quadratic in the torsion potentials. As for eqs. (\ref{e:3_2}) and (\ref{e:3_3}), the goal of a propagating description for torsion has been achieved: two second-order \emph{PDE}'s for both potentials have been obtained. To conclude, we write down the gauge transformations for the tensor potential
\begin{equation}
A_{\mu\nu}\to A'_{\mu\nu}=A_{\mu\nu}+\tilde{\nabla}_\mu Y_\nu-\tilde{\nabla}_\mu Y_n\;,
\end{equation}
by which, setting $Y$ such that $\tilde{\nabla}_\mu A'^{\mu\nu}=0$, it's easy to see that eq. (\ref{e:3_2}) rewrites as $\Delta_{\scriptscriptstyle{DR}}(A') = 0$, where $\Delta_{\scriptscriptstyle{DR}}$ is the de Rham operator. It is worth noting that, as far as eq. (\ref{e:3_3}) is concerned, it recasts a massless Klein-Gordon field equation is recovered, so that the potential $\phi$ can be considered as a geometrical manifestation of this field.

\subsection{Test-particle motion}

The problem of determining the equations of motion of a test particle is approached by several points of view \cite{ppptr,hjmn2}. Since torsion enters the expression of the covariant derivative of a vector, it affects motion: therefore the correct method is to perform the minimal substitution $(\nicefrac{d}{d\tau})\to(\nicefrac{\nabla}{d\tau})$. 

According to this rule, the motion equation in curved space is derived from that of special relativity $\nicefrac{du^{\mu}}{d\tau}=0$ ($u^\mu$ being the 4-velocity), for which $\nicefrac{\nabla u^{\alpha}}{d\tau}=0$ is obtained: this expression can be rewritten as
\begin{align}\label{e:5_3}
\tfrac{\nabla u^\rho}{d\tau}=
-\tilde{\Gamma}_{\mu\nu}^\rho u^{\mu}u^{\nu}- \nicefrac{2}{3}\;g^{\rho\sigma}
(g_{\mu\nu}\partial_\sigma\phi-g_{\mu\sigma}\partial_\nu\phi)u^{\mu}u^{\nu}\;.
\end{align}
This is the \emph{autoparallel equation}, which defines special curves in non-flat spaces, together with the geodesic equation: the latter is the shortest curve joining two points and autoparallels are those curves whose tangent vector is parallelly transported along it. The autoparallel curve is the simplest generalization of the flat-space motion equation, which is suitable to take into account torsion or other non-Riemannian quantities.

\paragraph{New action principle and non-holonomic map} This approach is proposed in \cite{klnrt2,klnrt1} and is based on the idea that it is possible to introduce a new action principle such that, starting from a modified action, autoparallels are obtained as the right trajectories. The key point is that a space-time with torsion, which can be obtained by a non-holonomic mapping from a flat space-time, 
is characterized by open (non-close) parallelograms; as a consequence, variations of test-particle trajectories cannot be performed keeping $\delta x^a(\tau)$ vanishing at endpoints. In fact, the variation images of $\delta x^a(\tau)$ under a non-holonomic mapping are generally not closed; this way, they can be chosen to be zero at the initial point but then they are non-vanishing at the final point. This behavior is due to torsion. 

\paragraph{Autoparallels from a modified action} Since the autoparallel motion can be derived from the energy-momentum-tensor ($\mathcal{T_{\mu\nu}}$) conservation law, we now give a possible modification of the test-particle action, such that this result could be partially obtained. To this end, we assume the test-particle action of the form
\begin{equation}
\mathcal{S}^{M}=\textstyle{\int}d\tau\,g_{\mu\nu} u^\mu u^\nu\,e^{-\nicefrac{\phi}{4}}\;.
\end{equation}
Taking into account the identification
\begin{equation}
\delta\mathcal{S}=\textstyle{\int}d^4x\sqrt{-g}\;
({}^{g}\mathcal{T}^{\mu\nu}\;\delta g_{\mu\nu}+{}^{\phi}\mathcal{T}\;\delta\phi)\;,
\end{equation}
we now calculate the action variations with respect to $g_{\mu\nu}$ and $\phi$, respectively:
\begin{equation}
\begin{split}\label{e:5_7}
{}^{g}\mathcal{T}^{\mu\nu}=\tfrac{\delta\mathcal{S}^{M}}{\delta g_{\mu\nu}}=
\textstyle{\int}\nicefrac{d\tau}{\sqrt{-g}}\;\,u^\mu u^\nu
e^{-\nicefrac{\phi}{4}}\;\delta(x-x_0)\;,\\
{}^{\phi}\mathcal{T}=\tfrac{\delta\mathcal{S}^{M}}{\delta\phi}=
-\nicefrac{1}{4}\textstyle{\int}\nicefrac{d\tau}{\sqrt{-g}}\;\,g_{\mu\nu}\,u^\mu u^\nu
e^{-\nicefrac{\phi}{4}}\;\delta(x-x_0)\;.
\end{split}
\end{equation}

Following \cite{hmmnd}, we consider the motion of a test particle, which negligibly perturbs the background geometry in which it lives, and start from the identity
\begin{equation}
(\sqrt{-g}\;{}^{g}\mathcal{T}^{\mu\nu})_{,\,\nu}=
\sqrt{-g}\;\;{}^{g}\mathcal{T}^{\mu\nu}_{;\,\nu}-
\sqrt{-g}\;\tilde{\Gamma}_{\rho\sigma}^{\mu}\;\,{}^{g}\mathcal{T}^{\rho\sigma}\;. 
\end{equation}
Let us now integrate the last expression over a volume $dV$, where the test-particle energy-momentum tensor is the only non-negligible one. Taking into account the conservation law
\begin{equation}
{}^{g}\mathcal{T}^{\mu\nu}_{\;;\,\nu}=
\nicefrac{8}{3}\;\partial^{\mu}\phi\;\;{}^{\phi}\mathcal{T}\;,
\end{equation}
and discarding all surface terms, we get
\begin{equation}
\tfrac{d}{u^0d\tau}\textstyle{\int}dV\sqrt{-g}\;\;{}^{g}\mathcal{T}^{\mu0}=\nicefrac{8}{3}\;\partial^{\mu}\phi\;\textstyle{\int}dV\sqrt{-g}\;\;
{}^{\phi}\mathcal{T}-\;\tilde{\Gamma}_{\rho\sigma}^{\mu}\;
\textstyle{\int}dV\sqrt{-g}\;\;{}^{g}\mathcal{T}^{\rho\sigma}\;.
\end{equation}
By (\ref{e:5_7}), this identity can be rewritten in the following form
\begin{equation}\label{e:5_8}
\tfrac{du^\rho}{d\tau}=-\tilde{\Gamma}_{\mu\nu}^{\rho}u^{\mu}u^{\nu}-
\nicefrac{2}{3}\;g^{\rho\sigma}(\partial_\sigma\phi)g_{\mu\nu}u^{\mu}u^{\nu}\;,
\end{equation}
and, if we multiply the lhs and the rhs of this equation by $u_\rho$, we obtain the identity
\begin{equation}\label{e:5_9}
0=u_\rho\;\partial^\rho\phi\;.
\end{equation}
Taking into account the autoparallel equation (\ref{e:5_3}), we immediate recognize that it matches the results (\ref{e:5_8}) and (\ref{e:5_9}), and Papapetrou motion is included as a special case.

\subsection{Non-relativistic limit and the role of the torsion potential}
On the basis of the minimal-substitution rule we have introduced, test particles follow autoparallel trajectories (\ref{e:5_3}). It is easy to see that the antisymmetric part of the torsion contribution vanishes; it only contributes as a source for the metric through (\ref{e:3_1}). In what follows, we will study the non-relativistic limit of autoparallels and in addition we will calculate the analogous of the geodesic deviation and we will see the role of torsion in the tidal forces.

\paragraph{Non-relativistic limit of autoparallels} To calculate the non-relativistic limit, the following hypotheses can be stated:\\ 
\;\emph{(i)} the \emph{3}-velocity is much smaller than $c$, so we can assume $u^i\simeq v^i$;\\
\;\emph{(ii)} the gravitational field and torsion potential $\phi$ are static and weak.

Since we want to keep only first order terms, by virtue of these assumptions, we will neglect all second-order terms in the quantities above. After some calculations, we obtain the autoparallel equation
\begin{align}\label{e:6.1_1}
\tfrac{dv_{i}}{dt}=-\nicefrac{\kappa}{2}\;\,\partial_{i}h_{00}-
\nicefrac{2}{3}\;\,\partial_{i}\phi.
\end{align}
where we have introduced the metric perturbation $h_{\mu\nu}=g_{\mu\nu}-\eta_{\mu\nu}$ ($\eta_{\mu\nu}$ being the Minkovsky metric). Now we recall that, in General Relativity (GR), we get the expression
\begin{align}       
\tfrac{dv_{i}}{dt}=-\nicefrac{\kappa}{2}\;\;\partial_{i}h_{00},
\end{align}
allowed us to identify $h_{00}$ with the gravitational potential $\Phi$,
\begin{align}
\nicefrac{\kappa}{2}\;\;h_{00}=\Phi.
\end{align}
As one can see from eq. (\ref{e:6.1_1}), the ``force'' due to the torsion potential is present in the same form of the gravitational field $h_{00}$; in addition, as for the order we are interested in, and reminding of the supposed field's static nature, eq. (\ref{e:3_3}) for the field $\phi$ reduces to
\begin{align}\label{e:6.1_2}
\Delta\phi(\textbf{x})=0, 
\end{align}
which recasts the gravitational field one
\begin{align}\label{e:6.1_3}
\Delta h_{00}(\textbf{x})=4\pi\rho.
\end{align}

\paragraph{Deviation of autoparallels} Since test particles move along autoparallels, we are able to calculate the relative acceleration between two such objects. Assuming two particles initially very 
close to each other, we obtain the expression   
\begin{equation}\label{e:6.2_1}
\tfrac{\tilde{\nabla}^2s^{\rho}}{d\tau^2}=
-\tilde{R}_{\mu\nu\sigma}^\rho\;s^\mu u^\nu u^\sigma+
-K_{\sigma\nu}^\rho(\tfrac{ds^{\nu}}{d\tau}\;u^{\sigma}+
\tfrac{ds^{\sigma}}{d\tau}\;u^\nu)-
(\tilde{\nabla}_{\mu}K_{\sigma\nu}^\rho)\;s^{\mu}u^{\sigma}u^{\nu}\;.
\end{equation}
Here $s^\mu$ is an infinitesimal vector representing the relative displacement between the two particles. This equation represents the generalization of the \emph{geodesic deviation} of standard GR to a theory with torsion. Once again, we note that the completely antisymmetric part of torsion contributes to the field equation only as a source.

In order to perform the non-relativistic analysis, we still keep the hypotheses \emph{(i)} and \emph{(ii)} above assuming now that velocity can be written as $\nicefrac{dx^\mu}{d\tau}\sim(1,0,0,0)$ and that particles accelerations are compared at the same time, \emph{i.e.}, $s^0=\nicefrac{ds^0}{d\tau}=0$.

Within this scheme, only terms containing $h_{\alpha\beta}$ or $\phi$ as factors multiplied times $s^i$ are non-negligible. Substituting the expression of the contortion tensor, eq. (\ref{e:6.2_1}) reduces to
\begin{align}
\tfrac{d^2s^i}{dt^2}\simeq
-\tilde{R}_{j00}^i \;s^{j}-\nicefrac{2}{3}\;\;\eta^{ij}\;s^{k}\;\partial_{kj}\phi\;,
\end{align}
this way, the tidal field becomes
\begin{align}
\mathcal{G}^i=-\tilde{R}_{j00}^i\;s^{j}-
\nicefrac{2}{3}\;\;\eta^{ij}\;s^{k}\;\partial_{k}\partial_{j}\phi\;.
\end{align}
From the non-relativistic limit of GR, we can identify
\begin{align}
\tilde{R}_{j00i}=\partial_{j}\partial_{i}\Phi\;,
\end{align}
where $\Phi$ is the gravitational potential. The final expression for the tidal 
field writes as follows:
\begin{align}\label{e:6.2_2}
\mathcal{G}_i=-s^j\;\partial_{i}\partial_{j}\Phi-
\nicefrac{2}{3}\;\;s^j\;\partial_{i}\partial_{j}\phi\;.
\end{align}
We can conclude that, in the non-relativistic limit, torsion produces a tidal-force effect analogous to that produced by the gravitational field.

It is worth noting that, since the fields $h_{00}$ and $\phi$ (in the non-relativistic limit) obey the Poisson \emph{PDE}'s (\ref{e:6.1_2}) and (\ref{e:6.1_3}) and enter eqs. (\ref{e:6.1_1}) and (\ref{e:6.2_2}) in the same way, it is impossible to distinguish the effect of the torsion field from that of the gravitational one, unless the source and the initial condition for the latter are know exactly; this fact, together with the small intensity of torsion forces, makes them even more difficult to be detected.

\section{Microscopic paradigm: Lorentz gauge theory}

Let $M^{4}$ be a \emph{4}-dimensional pseudo-Riemannian manifold, with a metric tensor $g_{\mu\nu}$, and $e$ a one-to-one map on it, $e:M^{4}\rightarrow TM^{4}_{x}$, which sends tensor fields on $M^{4}$ in tensor fields in the Minkowskian tangent space $TM^{4}_{x}$: the fields $e^{\phantom1a}_{\mu}$ (\emph{tetrads} or vierbein) are an orthonormal basis for the local Minkowskian space-time
\footnote{Latin indices from the beginning of the alphabet ($a=0,1,2,3$) transform under local Lorentz transformations.}. 
Given $\{e^{\phantom1a}_{\mu}\}$, the metric tensor $g_{\mu\nu}$ is uniquely determined, and all metric properties of the space-time are expressed by the tetrad field, accordingly, but the converse is not true: there are infinitely many choices of the local basis that reproduce the same metric tensor, because of the local Lorentz gauge invariance
\footnote{In fact, in the coordinate formalism, an infinitesimal diffeomorphism  and an infinitesimal Lorentz (isometric) rotation read
\begin{equation}\label{inf diff} 
x^{\mu}\rightarrow x^{\prime \mu}=x^{\mu}+\xi^{\mu}\left(x\right)\;,\qquad\quad
x^{\mu}\rightarrow x^{\prime \mu}=x^{\mu}+\epsilon^{\mu}_{\phantom1\nu}x^{\nu}\;,
\end{equation}
respectively, where $\xi^{\mu}\left(x\right)$ are four $C^{\infty}$ functions and $\epsilon^{\mu}_{\phantom1\nu}$ are the six infinitesimal rotational parameters. For local Lorentz transformations, where $\epsilon^{\mu}_{\phantom1\nu}\rightarrow\epsilon^{\mu}_{\phantom1\nu}(x)$, the second of eq. (\ref{inf diff}) can be easily reabsorbed into the first. In the non-coordinate formalism, it is possible to project the tensor field from the 4-dimensional manifold to the Minkowskian space-time, thus emphasizing the local Lorentz invariance of the scheme. Moreover, to assure that the derivative of a tensor field be invariant under local Lorentz transformations, the connection 1-forms $\omega^{a}_{\phantom1b}$ must be introduced: they defines the covariant exterior derivative operator $d^{(\omega)}$.\\
The connection 1-forms lead to the usual definition of the
curvature 2-form $R^{a}_{\phantom1b}$,
\begin{equation}
R^{a}_{\phantom1b}=d\omega^{a}_{\phantom1b}+\omega^{a}_{\phantom1c}\wedge\omega^{c}_{\phantom1b},
\end{equation}
which is the first Cartan structure equation. In this formalism, the action for GR consists of the lowest-order non-trivial scalar combination of the Riemann curvature 2-form and the tetrad fields, that is the Hilbert-Einstein action:
\begin{equation}\label{action for o}
\mathcal{S}(e,\omega)=\nicefrac{1}{4}\textstyle{\int}\epsilon_{a b c d}\,e^{a}\wedge e^{b}\wedge R^{c d}.
\end{equation}
Variation with respect to the connections leads to the second Cartan structure equation in the torsion-less case,
\begin{equation} \label{Cartan eq}
d e^{a}+\omega^{a}_{\phantom1b}\wedge e^{b}=0,
\end{equation}
while variation with respect to the tetrad leads to the following equations:
\begin{equation}
\epsilon_{a b c d}\,e^{b}\wedge R^{c d}=0,
\end{equation}
which, once the solution of the second Cartan structure equation (\ref{Cartan eq}) is considered, give the dynamical Einstein field equations.}
.

\subsection{LGT on flat space-time}
In flat space-time, the Riemann curvature tensor vanishes and, consequently, the usual spin connections $\omega^{a b}$ vanish too: this allows us to introduce Lorentz-valued connections as the gauge fields of pure local Lorentz transformations for spinors, and to define the tangent bundle
\footnote{Let $M^{4}$ be a \emph{4}-dimensional flat manifold: the metric tensor $g_{\mu\nu}$ reads
\begin{equation}\label{gen}
g_{\mu\nu}=\eta_{\alpha\beta}\tfrac{\partial x^{\alpha}}{\partial y^{\mu}}\tfrac{\partial x^{\beta}}{\partial y^{\nu}}=\eta_{\alpha\beta}e^{\alpha}_{\mu}e^{\beta}_{\nu},
\end{equation}
where $e^{\alpha}_{\mu}$ are bein vectors, $x^{\alpha}$ are Minkowskian coordinates, and $y^{\mu}$ are generalized coordinates. For an infinitesimal generic diffeomorphism
and for an infinitesimal local Lorentz transformation the behavior of a  vector field must be the same: from the comparison of the two transformation laws, the identification $\epsilon _{\alpha}^{\ \beta}\equiv \nicefrac{\partial \xi_{\alpha}(x^{\gamma})}{\partial x^{\beta}}$ is possible, where the isometry condition $\partial_{\beta}\xi_{\alpha}+\partial_{\alpha}\xi_{\beta}=0$
has to be taken into account in order to restore the proper number of degrees of freedom of Lorentz transformations, \emph{10}, out of that of generic diffeomorphisms, \emph{16}. The coordinate transformation that induces vanishing Christoffel symbols in the point $P$ is
\begin{equation}\label{cris}
y^{\alpha}_{P}=x^{\alpha}_{tb}+\tfrac{1}{2}\left[\Gamma^{\alpha}_{\beta\delta}\right]_{P}x^{\beta}_{tb}x^{\delta}_{tb},
\end{equation}
where $tb$ refers to the tangent bundle: the comparison with a generic diffeomorphism leads to the identification in the point $P$
\begin{equation}\label{map}
x^{\alpha}_{P}=x^{\alpha}_{tb}+\tfrac{1}{2}\left[\Gamma^{\alpha}_{\beta\delta}\right]_{P}x^{\beta}_{tb}x^{\delta}_{tb}-\xi^{\alpha},
\end{equation}
\emph{i.e.}, the coordinates of the tangent bundle are linked point by point to those of the Minkowskian space through the relation (\ref{cris}), and they differ for the presence of the infinitesimal displacement $\xi$. From now on, these coordinates will be referred to as $x^{a}$ (lower-case latin labels).},
where these transformations take place. Invariance under infinitesimal local Lorentz transformations
\begin{equation}
S\left(\Lambda\right)=1-\nicefrac{i}{4}\;\epsilon^{ab}(x)\Sigma_{ab}\;,
\end{equation}
is assured by the definition of covariant derivatives,
\begin{equation}
D_{a}\psi =e^{\mu}_{\phantom1a}D_{\mu}\psi=e^{\mu}_{\phantom1a}\left(\partial_{\mu}\psi-\nicefrac{i}{4}\;A_{\mu}^{b c}\Sigma_{b c}\psi\right),
\end{equation}
provided that the $\gamma$ matrices transform locally as vectors, and that the Lorentz gauge fields $A_{\mu}^{b c}$ transform as in the Yang-Mills scheme, for which a proper gauge-invariant Lagrangian has to be introduced. As a result, the interaction Lagrangian
\begin{equation}
\mathcal{L}_{int}=\tfrac{1}{8}\,e^{\mu} _{\phantom1a}\,\overline{\psi}\left\{\gamma^{a},\Sigma_{b c}\right\}\psi A^{b c}_{\mu}=-S_{b c}^{\mu}A^{b c}_{\mu}\;,\qquad\quad 
S^{a b}_{\mu}=-\nicefrac{1}{4}\;\epsilon^{a b}_{\phantom1\phantom1c d}e_{\mu}^{\phantom1c}j_{A}^d
\end{equation} 
illustrates that the spinor axial current $j_{A}^d=\overline{\psi}\gamma_5\gamma^d\psi$ both interacts with the gauge field and is the source of the gauge field itself. Field equations point up that the dynamics for a spinor field in an accelerated frame differs from the standard Dirac dynamics for the spinor-gauge field interaction term, \emph{i.e.}, spinor fields are not free fields any more.

\subsection{LGT on curved space-time}

\paragraph{Second-order approach} Generalizing the previous considerations on curved space-time, if anti-symmetric connections are hypothesized, the Lorentz gauge field can be identified with a suitable bein projection of the contortion field, \emph{i.e.}, $A_{ab\mu}\equiv -K_{\rho\sigma\mu}e^{\rho}_{a}e^{\sigma}_{b}$\,.\\
The comparison between local Lorentz transformations and gauge transformations allows one to obtain the expression for conserved quantities. This way, since the current density $J^{\mu}_{ab}\equiv\bar{\psi}_{r}\gamma^\mu\Sigma_{ab}^{rs}\psi_{s}$ admits the conservation law $D_{\mu}J^{\mu}_{ab}=0$, a conserved (gauge) charge\footnote{This quantity is a conserved one if one assumes that the fluxes through the boundaries of the space integration vanish.} can be defined
\begin{equation}\label{mah}
Q^{ab}=\textstyle{\int} d^{3}x J^{0 ab}=const.\;;
\end{equation}
on the other hand, the bein projection of the spin term of the angular momentum tensor $M^{\mu\nu}$, the conserved quantity for Lorentz transformations in flat space-time, reads
\begin{equation}\label{mahh}
M^{ab}=\textstyle{\int}  d^{3}x  \pi_{r}\Sigma^{ab}_{rs}\psi_{s}= const.\;,
\end{equation}
which coincides with (\ref{mah}), provided that $\pi_{r}$ is the  density of momentum conjugate to the field $\psi_{r}$, \emph{i.e.}, $\pi_{r}=\nicefrac{\partial L}{\partial \dot{\psi}_{r}}$. This identification is possible because of the definition of the parameter $\epsilon^{ab}$, which points up the remarkable features of local Lorentz transformations on the tangent bundle.

\paragraph{First-order approach} If one relaxes the torsion-less assumption, the second Cartan structure rewrites
\begin{equation}\label{Cartan eq general}
d e^{a}+\omega^{a}_{\phantom1b}\wedge e^{b}=T^{a}\;,
\end{equation}
where $T^{a}$ is the torsion 2-form; this equation is solved by the connections
\begin{equation}\label{connection}
\omega^a_{\phantom1b}=\widetilde{\omega}^a_{\phantom1b}+K^a_{\phantom1b}\;,
\end{equation} 
where $K^{a}_{\phantom1b}$ is the contortion 1-form, such that $T^{a}=K^{a}_{\phantom1b}\wedge e^{b}$, while $\widetilde{\omega}^a_{\phantom1b}$ are the usual Ricci spin connections. As a result, new 1-forms appear in the dynamics, which reestablish the proper degrees of freedom for the connections of the Lorentz group. In GR, nevertheless, these connections do not describe any physical field: after substituting the solution (\ref{connection}) of the structure equation into the Hilbert-Palatini action
\footnote{Let $\mathcal{S}\left(q_i,Q_j\right)$ be an action depending on two sets of dynamical variables, $q_i$ and $Q_j$. The solutions of the dynamical equations are extrema of the action with respect to both the two sets of variables: if the dynamical equations $\nicefrac{\partial\mathcal{S}}{\partial q_i}=0$ have a unique solution, $q_i^{(0)}\left(Q_j\right)$ for each choice of $Q_j$, then the extrema of the pullback $\mathcal{S}\left(q_i\left(Q_j\right),Q_j\right)$ of the action to the set of solution are precisely the extrema of the total total action $\mathcal{S}\left(q_i,Q_j\right)$.},
one finds that connections $K^a_{\phantom1b}$ appear only in a non-dynamical term, unless spinors are taken into account: in this case, the connections $K^a_{\phantom1b}$ become proportional to the spin density of the matter, thus giving rise to the Einstein-Cartan model, where the usual four-fermion term arises. 
For our purposes, we write the total connections as
\begin{equation}\label{connection-1}
C^a_{\phantom1b}=\omega^a_{\phantom1b}+A^a_{\phantom1b}, 
\end{equation}
where $\omega^a_{\phantom1b}$ are the standard connections of GR, and $A^a_{\phantom1b}$ are the connection 1-forms for local Lorentz transformations, whose appearence is connected with the presence of torsion, as it can be inferred from the comparison of (\ref{connection}) and (\ref{connection-1}). If the proper geometrical interpretation has to be attributed to the field $A$, the interaction term between the spin connections $\omega$ and the fields $A$
\begin{equation}\label{interacting term}
\mathcal{S}_{int}=2\textstyle{\int}\epsilon_{a b c d}\,e^{a}\wedge e^{b}\wedge\omega^{[c}_{\phantom1f}\wedge A^{f d]}
\end{equation}
has to be postulated. 

If fermion matter is absent, variation with respect to the connections $\omega$ gives, after standard calculations,
\begin{equation}\label{equation for omega}
d^{(\omega)}e^a=A^a_{\phantom1b}\wedge e^b,
\end{equation}
which admits the solution
\begin{equation}\label{solution vacuum}
\omega^a_{\phantom1b}=\widetilde{\omega}^a_{\phantom1b}+A^a_{\phantom1b},
\end{equation}
were $\widetilde{\omega}^a_{\phantom1b}$ are the usual Ricci connections: because of the analogy with the solution of the second Cartan structure equation (\ref{connection}), the connections $A$ can be identified with the 1-forms $K$. Since solution (\ref{solution vacuum}) is unique, the total action can be pulled back to the given solution to obtain the reduced action for the system
\footnote{Variation with respect to the gravitational field and connections of the Lorentz group leads to
\begin{equation}\label{pd}
\epsilon^a_{\phantom1b c d}\,e^{b}\wedge\widetilde{R}^{c d}=M^a+\epsilon^a_{\phantom1b c d}\,e^{b}\wedge\left(\widetilde{\omega}^{c}_{\phantom1f}+A^{c}_{\phantom1f}\right)\wedge A^{f d},\qquad
d^{(A)}\star F^{f d}=\epsilon_{a b c}^{\phantom1\phantom1\phantom1[d}\,e^a\wedge e^b\wedge\left(\omega^{c f]}+2A^{c f]}\right)\;,
\end{equation}
where $M^a$ is the energy-momentum 3-form of the field $A$, which can be explicitly obtained after variation of the Yang-Mills-like action with respect the gravitational 1-form.}.

If fermion matter is taken into account, variation of the total action with respect to the connections $w$ leads to
\begin{equation}\label{general equation for omega}
d^{(\omega)}e^a=A^a_{\phantom1b}\wedge e^b-\tfrac{1}{4}\epsilon^a_{\phantom1b c d}e^b\wedge e^c j_{(A)}^d,
\end{equation}
where the spinor axial current deeply modifies eq. (\ref{equation for omega}). Eq. (\ref{general equation for omega}) admits the unique solution
\begin{equation}
\omega^a_{\phantom1b}=\widetilde{\omega}^a_{\phantom1b}+A^a_{\phantom1b}+\tfrac{1}{4}\epsilon^a_{\phantom1b c d}e^c j_{(A)}^d,
\end{equation}
which can be inserted in the total action
\footnote{Variation with respect to the remaining fields leads a generalization of the dynamical equations (\ref{pd}).
Consequently, the density of spin of the fermion matter is present in the source term of the Yang-Mills equations for the Lorentz connection field, and the Einstein equations contain in the rhs not only the energy-momentum tensor of the matter, but also a four-fermion interacting term. The dynamical equations of spinors are formally the same as those of the Einstein-Cartan model with the adjoint of the interaction with the connections of the Lorentz group $A$.}.

\paragraph{Comparison} Since, in the first-order approach, the gravitational field plays the role of source for torsion, it should be compared with the ``current'' term of the second-order formalism. We will restrict our analysis to the linearized regime
\footnote{For small perturbations $h_{\mu\nu}$ of a flat-Minkowskian metric $\eta_{\mu\nu}$, $g_{\mu\nu}=\eta_{\mu\nu}+h_{\mu\nu}$, the  tetrad field rewrites as the sum of the Minkowskian bein projection $\delta^{a}_{\mu}$ and the infinitesimal perturbation $\zeta^{a}_{\mu}$, $e^{a}_{\mu}=\delta^{a}_{\mu}+\zeta^{a}_{\mu}$: the following identifications hold
\begin{equation}
\eta_{\mu\nu}=\delta_{\mu}^{a}\delta_{a\nu},\ \ h_{\mu\nu}=\delta_{a\mu}\zeta^{a}_{\nu}+\delta_{a\nu}\zeta^{a}_{\mu}.
\end{equation}
and the linearized Ricci connections $\omega^{ab}_{\ \ \mu}=e^{b\nu}\nabla_{\mu}e^{a}_{\nu}$ rewrite
\begin{equation}\label{corre1}
\omega^{ab}_{\ \ \mu}=\delta^{b\nu}\left(\partial_{\nu}\zeta^{a}_{\nu}-\tilde{\Gamma}(\zeta)^{\rho}_{\mu\nu}\delta^{b}_{\rho}\right),
\end{equation}
where $\tilde{\Gamma}(\zeta)^{\rho}_{\mu\nu}$ are the linearized Christoffel symbols.}
in the transverse-traceless (TT) gauge.

Because of the interaction term (\ref{interacting term}) postulated in the first-order approach, it is possible to solve the structure equation and to express connections as a sum of pure gravitational (Ricci) connections plus other contributions, both in absence and in presence of spinor matter. From the Einstein Lagrangian density for $g_{\mu\nu}$ in the TT gauge, 
\begin{equation}
\mathcal{L}=\left(\partial_{\rho}h_{\mu\nu}\right)\left(\partial^{\rho}h^{\mu\nu}\right)\;,
\end{equation}
the spin-current density  associated with the spin angular momentum operator $M^{\tau}_{\alpha\beta}$ can be evaluated for a Lorentz transformation of the metric. In fact, if we consider the transformation 
\begin{equation}
g_{\mu\nu}\rightarrow\tfrac{\partial x^{\rho'}}{\partial x^\mu}\tfrac{\partial x^{\sigma'}}{\partial x^\nu}\;g_{\rho'\sigma'}\;,
\end{equation}
where $x'^{\rho}=x^{\rho}+\epsilon^{\rho}_{\ \ \tau}x^{\tau}$, then the current reads
\begin{equation}\label{corre2}
M^{\tau}_{\alpha\beta}=\tfrac{\partial L}{\partial h_{\mu\nu,\tau}}\Sigma^{\rho\alpha\beta\sigma}_{\mu\nu}h_{\rho\sigma}=\left(\eta^{c\mu}\zeta^{\nu,\tau}_{c}+\eta^{c\nu}\zeta^{\mu,\tau}_{c}\right)\Sigma^{\rho\alpha\beta\sigma}_{\mu\nu}\left(\eta_{f\rho}\zeta^{f}_{\sigma}+\eta_{f\sigma}\zeta^{f}_{\rho}\right)\;,
\end{equation}
where $\Sigma^{\rho\alpha\beta\sigma}_{\mu\nu}=\eta^{\gamma[\alpha}\left(\delta^{\rho}_{\gamma}\delta^{\beta]}_{\mu}\delta^{\sigma}_{\nu}+\delta^{\rho}_{\mu}\delta^{\sigma}_{\gamma}\delta^{\beta]}_{\nu}\right)$.\\
The two quantities (\ref{corre1}) and (\ref{corre2}) do not coincide: in fact, (\ref{corre1}) is linear in the $\zeta$ terms, because the interaction term (\ref{interacting term}) is linear itself, while (\ref{corre2}) is second order in $\zeta$ by construction. As suggested by the comparison with gauge theories, and with (\ref{corre2}) in particular, the interaction term should be quadratic. In this case, however, it would be very difficult to split up the solution of the structure equation as the sum of the pure gravitational connections plus other contributions.

\section{Concluding remarks}
This paper is aimed at investigating the possibility to describe torsion as a propagating field, from both a macroscopic and microscopic point of view. 

In the fist case, we have exposed the formulation of a geometrical theory, which is able to predict propagating torsion. Starting from the Einstain-Cartan static theory, we introduce here two torsion potentials, by which we construct both the completely antisymmetric part of torsion field and the trace part. To determine the equation of motion of a test particle in presence of this new geometric quantity, we have established a principle of minimal substitution which implies that autoparallels are the right trajectories. Finally, we have analyzed the analogue of the geodesic equation for autoparallels and studied the non-relativistic limit of this deviation. Within this scheme, autoparallel deviation illustrates that the torsion potential $\phi$ enters the dynamics just the same way as the gravitational field $h_{00}$, thus letting us envisage an arduous experimental detection.

According to the different behaviors of vectors and spinors under local Lorentz transformations, a metric-independent Lorentz gauge field has been postulated, and its interaction with spinors has been analyzed. The mathematical identification of such a gauge field with a suitable bein projection of the contorsion field is possible through the structure equation, if a unique linear interaction term between gauge fields and spin connections is postulated. As a result, a Riemannian source for the Yang-Mills equations is induced. The real vacuum dynamics of the Lorentz gauge connection takes place on a Minkowski space only, when the Riemannian curvature and the spin currents provide negligible effects. In fact, it is the geometrical interpretation of the torsion field as a gauge field that generates the non-vanishing part of the Lorentz connection on flat space-time. The predictions of first- and second-order approaches are compared in the linearized regime. The two result do not match in this approximation, thus suggesting one to introduce a quadratic interaction term. Despite many formal differences from PGT, a pure contact interaction for spinor fields is recovered for vanishing Lorentz connections, for which the Cartan structure equation provides non-zero torsion even when gauge bosons are absent. From this point of view, PGT can be qualitatively interpreted as the first-order approximation of our scheme, when the carrier of the interaction is not observable, because of its feeble interaction \cite{lhc,obata}.


\end{document}